\documentstyle[12pt]{article}

\begin{document}
\baselineskip=1.5\baselineskip

\renewcommand{\thefootnote}{\fnsymbol{footnote}}

\setcounter{equation}{0}
\setcounter{section}{0}
\renewcommand{\thesection}{\arabic{section}}
\renewcommand{\theequation}{\thesection.\arabic{equation}}

\pagestyle{plain}
\begin{titlepage}

\hfill{SPhT/93-004({\bf Revised})}
\vfill
\begin{center}
{\large{\bf  {Dimensionality of spacetime as a gauge-invariance
parameter}
{in Yang-Mills calculations}}}\par
\end{center}
\vskip 1.5cm
\centerline{Rajesh R. Parwani\footnote{email : parwani@wasa.saclay.
cea.fr}}
\vskip 0.5cm
\centerline{Service de Physique Th{\'e}orique, CE-Saclay}
\centerline{ 91191 Gif-sur-Yvette, France.}
\vskip 1.0cm
\centerline{PACS 11.15Bt, 12.38Bx, 12.38Qk.}

\centerline{1 February 1993}
\centerline{ Revised {April 21, 1993}}


\vskip 1.5 cm
\centerline{\bf Abstract}
 It is described how the dimensionality of space-time may be used
  to check the gauge invariance of perturbative calculations
           in pure Yang-Mills ($YM$) theories. The idea is based on
the fact that pure $YM$ theory in two dimensions is
perturbatively free. Thus  gauge-invariant
quantities evaluated in a $D$ dimensional pure $YM$ theory should
vanish as $D$ goes to two.
The procedure and various subtleties in its application
are illustrated by
           examples drawn from quarkless $QCD$ at zero and
           nonzero temperature.
           The inclusion of quarks and the use of background
           field gauges is briefly discussed.\\

\end{titlepage}

\section{Introduction}
Perturbative calculations of gauge invariant quantities necessarily
 proceed in a gauge noninvariant manner due to the gauge-fixing
required in the Lagrangian. In order to verify the gauge-invariance
of the final result, and to check against possible errors,
computations are usually repeated for different choices of the
gauge-fixing or they are performed in a general
class of gauges labelled by an arbitrary gauge-fixing parameter.
In the latter case, one ascertains that the dependence on the
gauge-parameter drops out for physical quantities. For Yang-Mills
($YM$) theories, the complicated tensor structure of the vertices
makes calculations in a general gauge containing a gauge parameter
extremely tedious. In this paper I describe  how, for pure $YM$
theories, one may perform calculations in any particular
gauge with a convenient propagator (e.g. Feynman) and yet retain a
nontrivial check on the
gauge-invariance of the result.

The idea uses the fact that pure YM theory
in two dimensional space-time is
perturbatively free.  This is established by going to the
axial ($A_{1} = 0$) gauge
whence the gauge self-interactions vanish.
Since, by definition, gauge-{\it invariant}
quantities are independent of the choice of gauge-fixing, all
gauge-invariant quantities  in pure $YM_{2}$ theory must vanish.
The strategy to
use this fact for calculating physical quantities in some  $D_{0}$
dimensional space-time is as follows : perform the Lorentz algebra
and loop integrals for an arbitrary $D$ dimensions; then if the
quantity being calculated is truly gauge-invariant,  a
{\it necessary}
condition is that it should vanish at $D=2$. In this way  the
dimensionality of space-time is used as a gauge-invariance
 parameter.

As will be seen later, for all but one example in this paper the $D$
dependence of the Lorentz algebra gives the sole useful check on
gauge-invariance. However we will encounter an example of a
gauge-invariant quantity whose only $D$ dependence is in the loop
integral. In order to treat all possibilities in a unified manner
it is necessary to adopt prescriptions for defining  the
$D \rightarrow  2$ limit in the integrals. Integrals like those
 from zero-temperature
Feynman diagrams are defined in $D$ dimensions by analytic
continuation  \cite{UV,IR,Col,Muta} with the $D \rightarrow
2$ limit
taken after doing the integrals. For nonzero temperature integrals
containing Bose-Einstein factors, an infrared cutoff will be
imposed.

The reader is advised that it is { \it not} the aim of this paper
 to
provide, in a single attempt, a perturbative analysis of pure $YM$
theory for {\it all} $D_0 \ge D \ge 2 $ dimensions (if indeed such
a  thing  is possible), but rather to define a pragmatic procedure
that
connects correctly calculated gauge-invariant quantities near
$D = D_0$ to the value zero at $D =2$. The prescriptions are needed
for the loop integrals because even if the  gauge-invariant
quantity
is well defined near $D= D_0$, in the limit  $D \rightarrow 2$ one
will encounter infrared (IR) singularities symptomatic of
lower dimensional field  theories. Of course the prescriptions
mentioned
were chosen because they gave sensible results for the examples
considered. They remain to be checked in other cases as the author
has no general proof of their validity. Note that for zero
temperature
type of integrals dimensional continuation is being used here to
extrapolate
gauge invariant quantities from $D =D_0$ down to $D =2$, in
contrast to its
usual role of regulating ultraviolet (UV) and IR
\cite{UV,IR,Col,Muta}
singularities near $D =D_0$.

The $D \rightarrow 2$ check described above cannot be used for
gauge-invariant
quantities which are dimension specific. An example is the
 perturbative beta
function of $D_0 =4$ $YM$ theory which gives information about
the UV behaviour of Green's functions. Within mass-independent
renormalisation
schemes, the beta function is scheme-independent up
to second order and is manifestly gauge independent when minimal
subtraction
is used \cite{Col,Muta}. It is a dimension specific quantity
because it is
obtained from the residue of the pole, as $D \rightarrow 4$, of
the coupling
constant renormalisation factor. $YM$ theory is super
renormalisable for $D <
4$ and therefore lacks the conventional UV beta function. It
is thus
not apparent if one may sensibly
extrapolate the conventional beta function beyond an infinitesimal
 range
near $D=4$.
More examples of dimension specific
gauge invariant quantities may be found in $D_0 =3$ pure $YM$
theory with
an added  Chern-Simons term \cite{CS}. The Chern-Simons term is
specific
to odd dimensions and so here again one does not, in general,
expect
gauge-invariant quantities to vanish as $D \rightarrow 2$.

The nice thing about performing the Lorentz algebra in $D$
dimensions
(in addition to the integrals) is that
it takes almost no more effort than in doing it for the physical
 $D_{0}$
dimensions. The benefit, as mentioned above, is that the
$D$ parameter used in a simple gauge provides one with an
algebraically
efficient way of checking gauge invariance. Of course, one may
use the $D$
parameter in conjunction with  a
conventional
gauge parameter ($\alpha$) to give additional checks
and insight. The $D$ parameter is a book-keeping device
keeping track of the ``relevant'' $(D-2)$
pieces in a calculation while the $\alpha$ parameter prefaces the
``irrelevant'' pieces.

What about fermions? Clearly $QCD_{2}$ with fermions is a
nontrivial
theory \cite{GtH}. Fortunately, the contribution of fermions to
amplitudes can be kept
track of by using the usual trick of working with  an arbitrary
$N_{f}$
copies of them. A gauge-invariant quantity must be separately gauge
invariant in the $N_{f}=0$ and $N_{f} \neq 0$ sectors. In the
first sector,
the calculations may be performed as described above using the $D$
parameter to check gauge-invariance while the $N_{f} \neq 0$ sector
can be analysed separately.
 Usually diagrams with one or more
fermion lines are algebraically simpler to deal with than those
 with only gluon lines so the methodology described here is not
without promise.

The idea outlined in the preceding paragraphs will be exemplified
in this paper for  zero and nonzero
temperature ($T = 1/{\beta}$)  pure $YM$ theory
with gauge group $SU(N_{c})$ at $D_{0} =4$.
In Sect.(2) gluon-gluon scattering at zero temperature is
considered at tree level. This is a relatively simple example since
there are no loop integrals to complicate matters. The metric
used in Sect.(2)
is Minkowskian, diag($g_{\mu \nu}$)$={(1, -1,....,-1)}$.
In Sects.(3-5) the examples are at nonzero $T$ and the metric is
Euclidean, $g_{\mu \nu} =\delta_{\mu \nu}$ (for orientation to nonzero
temperature field theory see, for example, \cite{Ber,GPY}).
The measure for loop integrals in Sects.(3-5) is

\begin{equation}
\int [dq] \equiv T \sum_{q_0} {\mbox{$\displaystyle{\int{d^{(D-1)}q
 \over (2\pi)^{(D-1)}}}$}} \, , \label{meas}
\end{equation}
where the sum is over discrete Matsubara frequencies
\cite{Ber,GPY}, $q_0 =2 \pi
nT$ for gauge bosons and ghosts, $n \in \cal{Z}$.
For quantities which depend on the external momenta, an analytic
continuation
to Minkowski space is made as usual after the loop sums are done
\cite{GPY}. In Sect.(3) the one-loop gluon self-energy is
considered and the
two prescriptions for loop-integrals are introduced while in
Sect.(4) a
discussion is given of ``hard thermal loops'' and propagator
poles in $D$ dimensions. Sect.(5) considers the free energy of a
 gluon
plasma to third
order. The ``plasmon'' contribution in $D$ dimensions requires
the simultaneous
use of both prescriptions introduced in Sect.(3), therefore
providing a check on
their consistency. The conclusion is in Sect.(6) while the Appendix
contains some expressions and discussion mentioned in the main
text.

 The following gauges will be frequently
referred to throughout the paper :  the strict Coulomb
 gauge ($\xi =0$ limit of the $({\nabla}.\vec{A})^2 / 2 \xi$ gauge-
fixing), the $\alpha$-covariant
gauge with gauge-fixing term $ ({\partial}_{\mu} A_{\mu})^2 /
{2(\alpha  +1)}$
 and the Feynman gauge ($\alpha =0$). The Feynman rules, being
standard
\cite{Col,Muta,Ber,GPY}, will not be spelled out. $D$- vectors
will be denoted
by uppercase and have Greek indices, $Q_{\mu} = (q_{0} , \vec{q}
)$,
$q \equiv |\vec{q} |$, and the $(D-1)$
spatial components will be labelled by Roman letters ($i,j$).
Keep in
mind that in $D$ dimensions the coupling $g^2$ has a mass
dimension $(4-D)$.\\

\setcounter{equation}{0}
\section{Gluon-gluon scattering}
The scaterring amplitude $M(gg \to gg)$, for two gluons into two
gluons,
 involves
at lowest order four Feynman diagrams \cite{GG}. The first comes
 from
the order $g^2$ four-point vertex in the Lagrangian while the
other three are formed from
two three-point vertices tied by a propagator and represent the
usual $s,t $
and $u$ channel scatterings. The sum of the four amputated
Feynman diagrams
gives the tensor $T_{\mu \nu \sigma \tau}$, where the Lorentz
indices indicate the
external gluon legs. The gauge-invariant amplitude is then given by

\begin{equation}
M = T_{\mu \nu \sigma \tau} \epsilon_{1}^{\mu} \epsilon_{2}^{\nu}
\epsilon_{3}^{\sigma}
\epsilon_{4}^{\tau} \, . \label{M}
\end{equation}

Here $\epsilon_{(n)}^{\mu} \equiv \epsilon^{\mu}(\vec{k},\lambda_{(n)}) $
 represents the
polarisation vector for the $n$-th ($n=1,2,3,4$) gluon with
{ \it physical}
 polarisation
$\lambda_{(n)}$ and on-shell momentum $K^{2} = k_{0}^{2} -{\vec{k}}^2
= K^{\mu}
\epsilon_{\mu}(\vec{k},\lambda) =0$. In practice one usually needs the squared
amplitude
summed over initial and final spin (and colour) variables.
Choosing the basis
$\epsilon_{\mu}(\vec{k},\lambda) \equiv (0,\vec{\epsilon})$, one has the
transverse projection
operator

\begin{eqnarray}
P_{\mu \nu}(K) &=& \sum_{\lambda} \ \epsilon_{\mu}(\vec{k},\lambda)
 \epsilon_{\nu}(\vec{k},\lambda) = (\delta_{ij} - {k_{i}k_{j}
\over k^2})\delta_{\mu i} \delta_{\nu j} \, . \label{P}
\end{eqnarray}

When the relation (\ref{P}), which is true in any dimension, is
used to
evaluate $\sum_{\lambda} |M|^2$ in $D$ dimensions, factors of
$D$ will appear.
For example $g^{\mu \nu} P_{\mu \nu} = (D-2)$ and so in
 particular $P_{\mu \nu}
=0$ in two dimensions because then there are no transverse
states. From Ref.{\cite{ES}} one obtains

\begin{equation} {\displaystyle \Sigma_{\mbox{\small spin,colour}}}
|M(gg \to gg)|^2 = 4g^4 N_{c}^2
(N_{c}^2-1) (D-2)^2 \, \left[3 -{ut \over s^2} - {us \over t^2} -
{st \over u^2} \right] \, . \label{M2}
\end{equation}
For $D=4$ this reduces to earlier results \cite{GG} and it also
vanishes when
$D \rightarrow 2$ as desired. However there are two subtleties
which should be
noted. Firstly, since the on-shell gluons are massless, there are
kinematic singularities in (\ref{M2}) even for $D \ne 2$ : for example,
$s = (K_{1}^{\mu} + K_{2}^{\mu})^{2} = 0 $ when $\vec{k_{1}}$ is parallel
to $\vec{k_{2}}$. As $D \rightarrow 2$,  the Mandelstam
variables ($s,t,u$) vanish when the vectors
$(\vec{k_{2}}, \vec{k_{3}},
\vec{k_{4}} )$ are respectively in the same direction as $\vec{k_{1}}$.
Thus  the $D \rightarrow 2$ limit
of (\ref{M2}) is unambiguous only if the
kinematic singularities are regulated.
Secondly, if one also averages over initial spins in $D$
dimensions, then
(\ref{M2}) is divided by $(D-2)$ for each of the incoming lines.
This averaging is fine if one is working near $D=4$ say
\cite{ES}, but is clearly inadvisable
if one wants to  check gauge-invariance by the $D \rightarrow 2$
procedure : In the $D \to 2$ method one
should check gauge-invariant quantities before performing other
extraneous $D-$dependent operations.\\

\setcounter{equation}{0}
\section{Self-energy}
The self-energy by itself is not a gauge-invariant quantity.
However at nonzero temperature there
is a gauge-invariant piece of it which is easy to extract at low
orders. This is the inverse screening length for static electric
fields,
also called the electric mass, $m_{el}$. If $\delta^{ab}
\Pi_{\mu \nu} (k_0, \vec{k})$
is the gluon polarisation tensor at nonzero temperature, then at
lowest order one may  define

\begin{eqnarray}
m_{el}^{2} &\equiv& - \Pi_{00}(0, {\vec{k} \rightarrow 0}) \ .
 \label{me1}
\end{eqnarray}
\\
At $D_0 =4$, the order $(gT)^2$ result for (\ref{me1}) is well
known
\cite {GPY,Nad}. Remarkably it was found in Ref.\cite{Toi} that
the next term
of order $g^2 |\vec{k}|T$ in the low momentum expansion of
 $\Pi_{00} (0, \vec{k})$ at one-loop
is independent of the $\alpha$ parameter in the
$\alpha$-covariant gauge and also has the same value in the Coulomb
gauge, thus suggesting that even this term is gauge-invariant.
I repeat here the analysis of Ref.\cite{Toi} using the $D$
parameter.
In the $\alpha$-covariant gauge one finds
for the sum of one-loop gluonic
and ghost diagrams, the relevant object

\begin{eqnarray}
\Pi_{00}(0, \vec{k}) &=&  {g^{2} N_{c} \over 2} [ A_{0}(\vec{k}) +
\alpha A_{1}(\vec{k}) + {\alpha}^2 A_{2}(\vec{k})]  \; ,
\label{me2}
\end{eqnarray}

where

\begin{eqnarray}
A_{0}(\vec{k}) &=& \int [dq] {2(D-2)(2q_{0}^2 -Q^2) + 4k^2 \over
 Q^2 [q_{0}^2
+ (\vec{q}-\vec{k})^2] } \; , \label{me3} \\
&& \nonumber \\
A_{1}(\vec{k}) &=& \int [dq] {2[4(\vec{k}.\vec{q})^2 -2k^2 Q^2 +
2q_{0}^2k^2
]  \over Q^4 [q_{0}^2 + (\vec{q}-\vec{k})^2] } \; , \label{me4} \\
&& \nonumber \\
A_{2}(\vec{k}) &=& \int [dq] {q_{0}^2 k^4 \over Q^4 [q_{0}^2
+ (\vec{q}-\vec{k})^2]^2 } \; . \label{me5}
\end{eqnarray}
\\
The only difference between the integrands in eqns.(\ref{me2} -
\ref{me5})
and the expressions  studied in
\cite{Toi,KK} is the presence of the factor $(D-2)$, coming
from the  Lorentz algebra, in eq.(\ref{me3}). This
factor is invisible in \cite{Toi,KK} because they work with
$D=D_{0}=4$. From  the above expressions, one gets for the
electric mass squared at order $g^2$ :

\begin{equation}
m_{el}^2 = g^{2} N_{c} (D-2) \int [dq] {(2q^2 -Q^2)\over Q^4} \, .
\end{equation}

After performing the frequency sum and angular integrals one
obtains

\begin{equation}
m_{el}^2 = g^2 N_{c} (D-2) \ T^{(D-2)} \omega(D) \ [ 2J(D) -I(D)]
\, , \label{mD1}
\end{equation}

where
\begin{eqnarray}
{1 \over \omega(D)} &=& 2^{(D-2)} \  \pi^{({D-1 \over 2})} \
\Gamma\left({D-1 \over 2}\right)  \; ,  \label{ang} \\
I(D) &=& \int_{0}^{\infty} dx \ x^{(D-3)} \ n_x  \, ,\label{I} \\
J(D) &=& {1 \over 2} \int_{0}^{\infty} dx \ x^{(D-3)} \
[n_x  - x {d \over dx} n_x] \, , \label{J} \\
n_x &=& 1/(e^x -1) \, . \label{BEF}
\end{eqnarray}

Both of the integrals $I(D)$ and $J(D)$ are IR finite for $D > 3$.
The second term in $J(D)$
may be
integrated by parts, and the surface term dropped when $D >3$,
resulting in

\begin{equation}
J(D) = {1 \over 2} (D-1) I(D) \; \; \; , \, D > 3 \; .
\end{equation}
The integral $I$ can be written in terms of gamma and zeta
functions \cite{GR}
\begin{equation}
I(D) = \Gamma(D-2) \ \zeta(D-2) \; \; \; , \, D > 3 \; . \label{I2}
\end{equation}
Thus one may write eq.(\ref{mD1}) as
\begin{equation}
m_{el}^2 = g^2 N_{c} (D-2) \ T^{(D-2)} \omega(D) \ \Gamma(D-1)
 \zeta(D-2) \; \; \; , \,
 D>3 \, .  \label{mD2}
\end{equation}

The divergence as $D \rightarrow 3$ shows up in the zeta-function.
 In a consistent calculation at $D_0 =3$, the logarithmic
 divergence in the naive expression for $m_{el}$ will be cutoff by
 $g^2/T$ \cite{EDH}. Suppose one continues (\ref{mD2}) down to
$D=2$. Then the result vanishes because of
the $(D-2)$ Lorentz factor. However this may
be fortuitous as it is related to the possibility of
simplifying $J(D)$ (\ref{J})
 through an integration by parts, dropping a surface term, and
 getting a result
proportional to $I(D)$, so that the square brackets in (\ref{mD1})
has no net singularity at $D =2$. In more complicated examples
 one may not be so lucky. Therefore a prescription will now be
introduced to handle the IR singularities in integrals like
$I(D)$ and $J(D)$ above. It is simply this :
integrals with Bose-Einstein factors will be interpreted for
$D \le 3$ with an infrared cutoff $\lambda$:
\begin{equation}
\int_{0}^{\infty} \rightarrow \int_{\lambda}^{\infty} \, .
\label{pres1}
\end{equation}

That is, the lowest order electric mass is given in $D > 3$
 dimensions by the expressions (\ref{mD2}) and is defined,
for the purpose of this paper, by (\ref{mD1} - \ref{BEF},
\ref{pres1}) in $D \le 3$ dimensions. The cutoff in
 (\ref{pres1}) is left unspecified
since it is required here only to allow the limit
$D \rightarrow 2$ to
be taken with impunity. If one is really interested in the problem
in $D_0 \le 3$ dimensions then the cutoff must be determined self-
consistently. In this paper the interest is in gauge-invariant
quantities
near $D_0 =4$ and the prescription (\ref{pres1}) allows the
connection
to be made with the free theory at $D =2$. The prescription
(\ref{pres1})
will be tested in Sects.(4,5).

Now consider the order $|\vec{k}|T$ term in (\ref{me2}) for $D=4$.
 As discussed in
\cite{Toi}, this can only arise from the infrared region of the
integrals.
That is, it only arises from the $q_{0}=0$ part of the frequency
sum
(\ref{meas}) in (\ref{me3}, \ref{me4}). For the gauge-fixing
dependent
piece (\ref{me4}), the zero
mode contains pieces exactly of order $|\vec{k}|T$ but the net
contribution vanishes after the elementary integrals are
done \cite{Toi}. The zero mode in the $\alpha$ independent piece
(\ref{me3})
contributes
\begin{eqnarray}
&&T \int {d^{(D-1)} q \over {(2 \pi)^{(D-1)} }} \left[ {-2(D-2)
\over (\vec{q} -\vec{k})^2} +
{4k^2 \over q^2 (\vec{q}-\vec{k})^2}\right] \; . \label{kT1}
\end{eqnarray}

The first term in (\ref{kT1}) vanishes by dimensional
regularisation. The
second gives, at $D=4$, the  contribution proportional to
$|\vec{k}|T$
found in \cite{Toi}. In $D$ dimensions this last piece has no
$(D-2)$
factor from the Lorentz algebra but the integral is highly singular
for $D \le 3$ even when $ \vec{k} \ne 0$. As the integral is
similar to
that occurring in zero temperature field-theory (indeed (\ref{kT1})
is a contribution in the effective $(D-1)$ dimensional Euclidean
field
theory which represents the far infrared,
or infinite temperature, limit of the $D$ dimensional finite
temperature field theory \cite{GPY,JT}.) it is natural to use
dimensional
continuation methods \cite{UV,IR,Col,Muta} for its evaluation.
A standard calculation of (\ref{kT1}) yields,

\begin{equation}
T \left({D-2 \over 4} \right) { k^{(D-3)}  \over
{(4 \sqrt{\pi})^{(D-4)}}} { 1 \over \Gamma(D/2) \cos(\pi D/2)} \,
. \label{kT2}
\end{equation}

Amazingly, $D=2$ is the only positive value of $D$ for which
(\ref{kT2}) vanishes. Thus the $(kT)$ term in (\ref{kT1}) at
  $D=4$ {\it does} satisfy the
 necessary condition  for gauge-invariance once the integral is
defined  by dimensional
continuation for the $D \rightarrow 2$ limit. Of course the above
analysis  does not explain {\it why} the $kT$ term is
gauge-invariant. In Ref.\cite{Toi}
it was related to a higher order term in the free energy but its
direct
physical significance is unclear to the present author. It might
be
interesting also to have a general proof for the gauge-invariance
of the $kT$ term using, for example, the techniques of
Ref.\cite{LR}.

The use of dimensional continuation to evaluate zero temperature
type integrals is the second prescription that will be used in this
paper. Another example of its use will be given in Sect.(5).
Here it  is noted that
with the replacement $(D-1) \to D$, the second integral  in
(\ref{kT1}) occurs  in the zero temperature
 self-energy in $D$ dimensions.
The zero-temperature self-energy thus diverges when
$D \rightarrow 2$
(i.e. $D \to 3$ in (\ref{kT2})) but this is not worrisome since
the  self-energy is a gauge-dependent object. \\

\setcounter{equation}{0}
\section{`Hard Thermal Loops' and propagator poles}
For $QCD_4$, at nonzero temperature, there are  an infinite number
of bare loop diagrams which are as large as the tree amplitudes
when
the momentum entering the external legs is soft ($ \sim gT$) and
the
internal loop momentum is hard $(\sim T)$. These ``hard thermal
loops'' (HTL)
occur only at one-loop and have been extensively analysed
by Braaten and Pisarski \cite{BP} and Frenkel and Taylor
\cite{FT}.  The HTL's exist for amplitudes when all the $N \geq 2$
 external lines are gluons or when one pair is fermionic and the
 other $(N-2)$ are gluons.
By explicit calculations \cite{BP,FT},
the HTL's were found to be the same in Coulomb,
$\alpha$-covariant and axial gauges. General proofs of
gauge-fixing  independence may be constructed \cite{KKR}.
A gauge-invariant generating functional for the HTL's that was
constructed by Taylor and Wong  has  been cast into myriad
forms \cite {TW}. In some
recent work, Blaizot and Iancu \cite{BI} have rederived the
results of
\cite{BP,FT,TW} by analysing the kinetic equations obtained
through a
self-consistent truncation of the Schwinger-Dyson equations for
sources and fields at finite temperature.

 From the expressions contained in  \cite{BP} or \cite{TW,BI} one
sees that
the $N_f=0$ sector of the $N$-gluon HTL contains an overall
factor of
$(D-2)$ when the Lorentz algebra is done in $D$ dimensions.
Even the HTL's with external quark lines are seen to be
proportional
to $(D-2)$. As noted in the above papers, this is because
the HTL's, which are the
leading high temperature (and essentially classical) parts of
 the one loop
diagrams, receive contributions only from the $(D-2)$ physical
transverse gluon degrees of freedom.

To consider the pure gluonic HTL's  in $D$ dimensions (the $N_f
\neq 0$ sector is not of interest here), the $D$ dependence of
the integrals
must also be taken into account (see also Frenkel and Taylor
 \cite{TW}).  For the purpose of power
counting it is convenient to  introduce the
dimensionless coupling $g_0$ in $D$ dimensions through the
relation $g^2 =
g_{0}^{2}T^{(4-D)}$, where the temperature has been  chosen as
the mass scale
since that is the natural parameter in the problem. Now a hard
momenta
is of order $T$ while soft refers to $\sim g_0 T$. With this
notation one can repeat all the relevant analysis of
\cite{BP,FT,TW,BI} and show
that it remains valid for $D > 3$ dimensions.
However naive power counting suggests that
for $D \le 3$ dimensions {\it soft} thermal loops (loop momenta
$\sim g_0 T$)
are no longer suppresed relative to HTL's.
This is related to the occurrence of IR divergences; for example,
the static
limit of the HTL in the gluon self-energy \cite{KW} is simply
the  electric
mass squared (\ref{me1}) which was noted in the last section to
diverge in the
naive $D \rightarrow 3$ limit. Therefore, just as in the case of
$m_{el} $ ,
for the purpose of taking the $D \rightarrow 2$ limit, HTL's are
defined
in this paper for $D \le 3$ with the infrared cutoff
(\ref{pres1}).  Then they vanish as $D \rightarrow 2$ simply
because of the Lorentz  algebraic factor.

Just as at zero temperature, the physical poles of the propagator
at non-zero  temperature are gauge invariant \cite{KKR}.
At nonzero temperature, the real part of the gauge propagator
 pole at zero external three momentum
defines the induced thermal masses for the gluons and for $D_0 =4$
 the  leading ($ \sim gT$) result is easily obtained at one-loop
\cite{GPY}. When using the $D$ parameter, the thermal mass will
vanish near $D =2$ as  $\sim \sqrt{(D-2)}$, just like the
electric mass (\ref{me1}), when the prescription (\ref{pres1}) is
adopted. The imaginary part (at $D_0 =4$) turns
out to be of subleading order ($g^2T$) and a practical
consistent calculation
requires the Braaten-Pisarski \cite{BP} resummation using
propagators and vertices dressed with HTL's. If the calculation
of the imaginary part is done in $D$ dimensions there will be
three sources of $D$ dependence : from the HTL's in the effective
propagators and vertices, from the Lorentz algebra of the
dressed diagrams, and from the loop integral of the dressed
diagrams. It
would be interesting to see how the $D \rightarrow 2$ limit
looks like
in this case but this will not be attempted here because the
analysis
is tedious. In the next section an example will be considered which
also involves a resummation but is easier to analyse.\\

\setcounter{equation}{0}
\section{Free energy}
The free-energy is physical quantity equal to the negative of
the pressure and  is directly obtainable by
calculating bubble diagrams in perturbation theory \cite{Kap}.
 Since it is physical,
it must be gauge-invariant. In the Feynman gauge, the ideal gas
pressure ($P_0$) of gluons is given by \cite{Ber,GPY}
\begin{eqnarray}
{ P_{0} V \over T } &=&  (N_{c}^2 -1) \  \ln \left\{ \left[
Det (-\partial ^{2}  \delta_{\mu \nu}) \right]^{-{1 \over 2}} .
 Det( -\partial ^2) \right\}  \label{det} \\
&=&  (D-2) (N_{c}^{2} -1) \ \ln [Det(-\partial ^{2} )]^{-{1
\over 2}} \, ,
 \label{IG}
\end{eqnarray}
where $V$ is the volume. The first determinant in (\ref{det})
 is the  contribution of gluons
while the second determinant is the ghost contribution. The first
two terms in (\ref{IG}) count the number of physical degrees of
freedom.
The remaining expression in (\ref{IG}) may be evaluated
(see appendix)
to yield the free gluonic pressure in $D$ dimensions,
\begin{equation}
P_{0} = (D-2) (N_{c}^{2}-1) \ T^{D} \ {\pi}^{-{D \over 2}} \
\Gamma(D/2)
\zeta(D) \; \; \; ,\  D > 1 \, . \label{IG2}
\end{equation}
The result is positive for $D > 2$ and vanishes smoothly in the
limit
$D \rightarrow 2$. The first singularity
appears in the zeta-function at $D =1$ when field theory
collapses to
quantum mechanics.

Consider next the order
$g^2$ correction to the ideal gas pressure, $P_2$. In the Feynman
gauge one obtains after some  algebra,

\begin{eqnarray}
P_2 &=& g^2 N_c (N_{c}^2 -1) \left[\int {[dq] \over Q^2}\right]^2
\left\{ -{ 1 \over 2} ({1 \over 2}) + {1 \over 8} [2D(1-D)] +
{1 \over 12} [9(D-1)] \right\}  \nonumber \\
& & \label{P21} \\
& = & -\left({{D-2} \over 2}\right)^2  g^{2} N_{c}(N_{c}^{2} -1)
\left[\int
{[dq] \over Q^2}\right]^2   \label{P22} \\
& = & -\left({{D-2} \over 2}\right)^2  g^{2} N_{c}(N_{c}^{2} -1) \
T^{(2D-4)} \ {\omega}^{2}(D) \  I^{2}(D) \; .   \label{P23}
\end{eqnarray}
\\
The terms within brackets in (\ref{P21}) come
respectively from the two-loop bubble
diagrams with
one, two and three  gluon propagators. Shown explicitly in front
of each contribution are the symmetry factors and the minus sign
 for the
ghost loop. The functions $\omega(D)$ and $I(D)$ in (\ref{P23})
are those
defined earlier in eqns.(\ref{ang}, \ref{I}). When $D>3$ one may
also use
eq.(\ref{I2}) and at $D=4$ one recovers a known
result \cite{Kap}. For $D <3$ the prescription (\ref{pres1}) is
again to be
used for the integral $I(D)$. Then the  net result in (\ref{P23})
vanishes for $D=2$ as required for a gauge invariant quantity.
The main point here is that if one had made errors (for example in
the symmetry factors in (\ref{P21})), these would likely have
shown up
in the nonvanishing
of the net result at $D=2$. A similar calculation
in an $\alpha$-covariant gauge for the purposes of checking
algebra
is far more tedious, especially for the diagram
with three gluon lines. The complexity of the algebra in an
{$\alpha$}-gauge in fact increases the sources of possible errors
at intermediate
steps. As a curiosity, it might interest the reader to note that
nevertheless the result (\ref{P22}) can
also be established in an $\alpha$-covariant gauge
{\it before} doing any explicit
integrals, albeit with greater algebraic effort,
the $\alpha$ dependence cancelling in the sum of
diagrams as required (see appendix).

The next correction to the pressure in four dimensions is of
order $g^3$.
This ``plasmon'' correction is a nonperturbative contribution and
it was
computed in $QCD$ by Kapusta \cite{Kap,PlasC}. It is obtained by
summing an infinite class of IR divergent diagrams, formed by
adding two or more self-energy subdiagrams along the gluon line
of the
one-loop bubble diagram.
The leading correction ($\sim g^3$) is due to the electric mass,
$\Pi_{00}(0,\vec{k} \to 0)$. Summing the electric mass insertions
 in $D$ dimensions
gives

\begin{equation}
 P_3 = -{ (N_{c}^2 -1) \over 2 \beta}  \int {d^{(D-1)} q \over
{(2 \pi)^{(D-1)} }}
\left[  \ln (1 + {m_{el}^2 \over q^2}) - {m_{el}^2 \over
q^2}\right] \, . \label{P31}
\end{equation}

The above expression is well defined for $D >3 $ with $m_{el}$
given by (\ref{mD2}).
The loop integral may be evaluated using zero temperature
techniques (see appendix)
to give
\begin{equation}
P_3 = { (N_{c}^2 -1) \over 2 \beta} { \Gamma({1-D \over 2})
(m_{el}^2)^{D-1 \over 2} \over (4 \pi)^{D-1 \over 2}} \; \; \;
, \ D > 3 \, . \label{P32}
\end{equation}

Since $m_{el} \sim g$, the result (\ref{P32})
is subleading, when $D >3$, to the order $g^2$ contribution
$P_2$ given
by eq.(\ref{P23}).
Also note that (\ref{P32}) is positive for $ 3<D <5$ so that it
opposes  $P_2$ in that range.
In order to apply the $D \to 2$ check on (\ref{P31}) we need
to use both of
the prescriptions introduced earlier. Firstly, for $D <3$ the
electric mass $m_{el}$ is defined
by the cutoff prescription (\ref{mD1}- \ref{BEF}, \ref{pres1}).
Secondly, the loop integral
in (\ref{P31}) is IR divergent for $D<3$ and so it is defined by
the analytic
continuation prescription (the IR divergence coincides with a
physical effect : the magnetic contribution may no longer be
subleading (see appendix)). Thus one takes
the $D \to 2$ limit in (\ref{P32}) with $m_{el}$ defined by
(\ref{mD1}, \ref{pres1}).
Since $ m_{el} \sim \sqrt{(D-2)}$, therefore when $D \to 2$,
$P_3$ vanishes as
$ \sim (D-2)^{\gamma}$ with $\gamma = {1 \over 2} + {(D-2) \over
2}$. The $D$ dependent exponent
$\gamma$ is another sign of the nonperturbative nature of the
plasmon term. The nontrivial
point here is that the loop integral has not introduced any
adverse powers
of $(D-2)$ which would have had a disastrous effect for the
 $D \to 2$ limit. This
example shows a successful cohabitation of the two IR
prescriptions
that were introduced  for defining the $D \to 2$ limit.

\section{Conclusion}
The dimensionality of spacetime ($D$) has been
proposed and illustrated
as a possibly  efficient and beneficial way to check
gauge-invariance in pure {YM}
theories. Gauge invariant quantities which are not dimension
specific should vanish as $D \to 2$.
The
converse is not necessarily true. For example, any quantity,
even if gauge
{\it variant}, when calculated in the axial gauge should vanish
as $D \to 2$ due to the free nature of pure $YM_2$.

Although in most of the examples it was the Lorentz algebra
which contained the
useful $D$-dependent information, the procedure required the
 use of two
{\it ad hoc} prescriptions to define the $D \to 2$ limit in
loop integrals.
Zero-temperature-type integrals were defined by analytic
continuation while
integrals containing a Bose-Einstein factor were cut off by an
infrared regulator.
In the examples considered the prescriptions allowed one to
extrapolate gauge-invariant
 quantities  calculated near $D =D_0 =4$ down to $D =2$ in the
required manner.
Instead of the two prescriptions, one might try the
following single
condition : analytically continued gauge invariant quantities
in pure $YM$
theory should be nondiverging at $D=2$. A relook at the
examples shows that
this also provides a nontrivial check.
In the absence of an {\it a priori} justification of the
prescriptions,  one is actually checking both the IR prescriptions
and the gauge-invariance. Still, the analysis of gauge-invariant
structures for a variable $D$ appears instructive and one might
want to consider more examples and at higher order. It might also
be interesting to explore the $D \to 2$ procedure for
gauge-invariant
quantities correctly evaluated near
$D_0 =3$ \cite{EDH,JT} .

Fermions can be accomodated by using the number of flavours,
$N_{f}$, as a
parameter. The $N_f \neq 0$  part of any gauge-invariant
quantity must be invariant by itself. At low orders in
perturbation theory,  one may even entertain the notion of
calculating the $N_f =0$ and $N_f \neq 0$ sectors with different
 gauge-fixing. For example, the pure glue part can
be calculated in the Feynman-$D$ gauge while the $N_f \neq 0$
part can be calculated in the $\alpha$-gauge to check
 gauge-invariance. Whether such
hybrid calculations are useful or practical should be decided
on a case by
case basis. Likewise, scalars can be coupled by taking $N_s$
copies of them.

Finally some comment on the background field gauge \cite{Dew}.
This is one way of
calculating in quantum field theory while keeping classical gauge
invariance at every step. The gauge-invariance here is with
respect to the background field $B_{\mu}$ which is introduced for
this purpose and gives no information about the physical
gauge-invariance
of any quantity calculated. In particular, the quantum part of
 the action
must still be gauge fixed. Thus even here one might use the $D$
 parameter
without redundancy.

{\noindent{\bf Acknowledgements}}\\
I thank J.P. Blaizot, C. Corian{\'o}, A.S. Goldhaber, E. Iancu,
H. Osborn, R.D. Pisarski and J.C. Taylor for very helpful
discussions. I also acknowledge stimulating and hospitable visits
to DAMTP-Cambridge and Martignano-Italy
during the course of this work.\\

\renewcommand{\theequation}{A.\arabic{equation}}
\setcounter{equation}{0}

\noindent{{\bf Appendix}}\\
1.Some  formulae are collected here for ease of reference.

a)The bosonic ($q_0 =2 \pi n T$) sums needed in Sect.(3) are
\cite{GPY}
\begin{eqnarray}
T \sum_{q_0} {1 \over Q^2} &=&  {n(q) \over q} + {1 \over 2q} \,
, \label{S1} \\
T \sum_{q_0} {1 \over Q^4} &=& {n(q) \over 2q^3 }- {1 \over 2q^2 }
{ d n(q) \over dq} +  { 1 \over 4 q^3} \; , \label{S2}
\end{eqnarray}
where $n(q) = (\exp{\beta q} -1)^{-1}$ is the Bose-Einstein
factor. The last terms in the above sums are temperature
independent and drop when dimensional
regularisation is used for the $ \vec{q}-$integrals .

b)The angular  integrals for $(D-1)$ dimensional Euclidean space
have been defined by \cite{Muta}
\begin{equation}
\omega(D) = \int {d \Omega_{D-1} \over (2 \pi)^{(D-1)}} \equiv
\left[ 2^{(D-2)} \ \Gamma \left({D-1 \over 2}\right) \
\pi^{(D-1) \over 2}
 \right]^{-1} \, .
\end{equation}

c)The zero temperature integrals in Sects.(3,5) are evaluated
using \cite{UV,IR,Col,Muta},
\begin{eqnarray}
\mbox{$\displaystyle{\int {d^{s}q \over (2\pi)^{s}}}$} {1 \over Q^2 (Q + K)^2}
&=& (4 \pi)^{-{s \over 2}} \
(K^2)^{(s-4)
\over 2} \ {\Gamma(2 -{s \over 2}) \Gamma^{2}({s \over 2} -1)
\over \Gamma(s-2)} \; \; , \\
\nonumber \\
\mbox{$\displaystyle{\int {d^{s}q \over (2\pi)^{s}}}$}
 {1 \over Q^2 + M^2 } &=& (4 \pi)^{-{s \over 2}} \
(M^2)^{(s-2) \over 2}
\ \Gamma(1- {s \over 2}) \, .
\end{eqnarray}

d)Expressions containing gamma-functions can be simplified with
the following
very useful identities \cite{GR}
\begin{eqnarray}
\Gamma(1+z) &=& z \ \Gamma(z) \; ,\\
\Gamma(z) \ \Gamma(1-z) &=& {\pi \over \sin{\pi z}} \; , \\
\sqrt{\pi} \ \Gamma(2z) &=& 2^{(2z-1)} \ \Gamma(z) \
\Gamma(z + {1 \over 2}) \; .
\end{eqnarray}
\\
\noindent{2. The one-loop gluon self energy is given by}
\begin{equation}
\Pi_{\mu \nu}^{a b}(K) = \int [dq] \{ L_{\mu \nu}^{ab}(K,Q) +
{1 \over 2}
M_{\mu \nu}^{a b}(K,Q) + {1 \over 2} N_{\mu \nu}^{ab}(K,Q) \} \;
, \label{Se}
\end{equation}
where $({\mu \nu})$ are the Lorentz indices and $(ab)$ the group
indices. The symmetry
factors have been explicitly displayed. $L$ is the ghost loop
 contribution
($-1$ factor included), $M$ the
tadpole diagram and $N$ is due to the tri-gluon coupling.
Expressions for $L,M$ and $N$
in $D$ dimensions may be found, for example, in \cite{Muta}.
The complete result
at zero temperature, in an $\alpha$-covariant gauge, with the
integrals done, may also
be found in \cite{Muta}. For the Landau gauge ($\alpha = -1$) the
expression is contained in Ref.\cite{DLM} which
studies $QCD$ in $2 +\epsilon$ ($\epsilon \ll 1$) dimensions and
also notes the divergence
of the self-energy as $\epsilon \to 0$.\\

\newpage
\noindent{3. The Free energy.}

a)The contribution of each massless, bosonic degree of freedom
to the ideal
gas pressure is
\begin{eqnarray}
P_{0}^{b} &=& -{1 \over 2 V \beta} \ln Det (- \partial^{2} ) \\
&=& -{1 \over 2} \int [dq] \ln (Q^2) \\
&=& -T \mbox{$\displaystyle{\int{d^{(D-1)}q \over (2\pi)^{(D-1)}}}$}
 \ln (1- e^{- \beta q}) \\
&=& -T^{D} \ \omega(D) \int_{0}^{\infty} dx x^{(D-2)}
\ln(1- e^{-x}) \\
&=& T^{D} \ \omega(D) \ \sum_{p= 1}^{\infty} \ {1 \over p}
\int_{0}^{\infty}
dx x^{(D-2)} e^{-p x} \\
&=& T^{D} \ \omega(D) \Gamma(D-1) \zeta(D) \\
&=& T^{D} \ \pi^{-{D \over 2}} \ \zeta(D) \Gamma(D/2)  \, .
\label{Pb}
\end{eqnarray}
The determinant above is evaluated with the required
periodic boundary conditions \cite{Ber,GPY}.
In the second line a $T$-independent piece was
dropped. The interchange, in order, of the integration and power
series
summation is justified for $D >1$. Final simplification is
achieved using the
definition of the gamma and zeta functions \cite{GR} and the use
of formulae
in Note(1) of this appendix. In passing it is noted that for
massless
fermions at zero chemical potential, each modes contribution to
the ideal
pressure will turn out to be eq.(\ref{Pb}) multiplied by a
statistical factor
$(1-2^{(1-D)})$. For the case of massive particles, nonzero
chemical
potentials and background fields, see \cite{Wel}.

b)For the calculation of the order $g^2$ contribution to the
pressure,
one can save some effort and reduce errors by proceeding as follows
\begin{equation}
P_2 = \int [dk] \int [dq] D(K) \{ {1 \over 2} L + {1 \over 8} M +
{1 \over 12} N \} \, .
\end{equation}
That is, compute the expression in curly brackets first.
Here $D(K) = {\delta^{ab} \over K^2} ( \delta{\mu \nu} + \alpha
{K^{\mu} K^{\nu}
 \over K^2})$ is the free propagator in the $\alpha$-covariant gauge
and $L,M$ and
$N$ are the $\alpha$ dependent tensors used in eq.(\ref{Se}) above.
The $\alpha$ dependent pieces
of $P_2$ cancel only after frequent use of the identity
$2K.Q = (K+Q)^2 -K^2 -Q^2$, changes of sum-integration variables,
and
shifts of sum-integration variables
(assumed valid), to obtain the final answer displayed in
(\ref{P21}). Expressions similar to (\ref{P21}) in the background
Feynman gauge may be found in \cite{SH}.

c)The plasmon contribution in four dimensions has been calculated
by Kapusta \cite{Kap} in the Feynman gauge. Here I sketch the $D$
dimensional analog in the Coulomb gauge, using the notation of
Toimela \cite{PlasC}.
One begins with
\begin{eqnarray}
P_{plas} &=&{1 \over 2 \beta} \int [dq] \sum_{p=2}^{\infty} \
{1 \over p} Tr (-D^{c} \Pi)^{p} \\
&=&  {(N_{c}^2 -1) \over 2 \beta} \int [dq] \sum_{p=2}^{\infty}
{(-1)^{p} \over p} \left[ \left(F \over q^2\right)^p +
(D-2) \left(G \over Q^2 \right)^p
\right]
\end{eqnarray}
In the above $D^{c}$ is the free propagator in the (strict)
Coulomb gauge
\begin{equation}
D^{c}_{\mu \nu} =  { \delta_{\mu 0} \delta_{\nu 0} \over k^2} +
{ \delta_{\mu i}
\delta_{\nu j} \over K^2} \left( \delta_{ij} -{k_i k_j \over k^2}
\right)  \, ,
\end{equation}
$\Pi$ is the one-loop self energy, $F \equiv \Pi_{00}$, and
$G$ is the transverse part of $\Pi_{ij} $ :\\
$(D-2)G = \Pi_{ij} ( \delta_{ij} -q_{i}q_{j}/q^2)$, with sum over
repeated indices.

In order to obtain the leading  plasmon-like ($ > {g^4}$)
contribution from $P_{plas}$,
one need only  look at the infrared region which lies in $q_0$
sector.
Now, in four dimensions we have $F \sim g^2 T^2$ and
$G \sim g^2 k T$. In $D$ dimensions near $D=4$ one therefore
expects $F \sim
g^2 T^{(D-2)}$ and $G \sim g^2 T k^{(D-3)}$. Using the
dimensionless
coupling $g_0$ defined by $g^2 =g_{0}^2 T^{(4-D)}$, and assuming
$g_0 \ll 1$,
consider the
contribution of soft ($\sim g_0 T$) loop momenta to the $p$-th
term in
$P_{plas}$. The electric ($F$ type) contribution will be
\begin{equation}
\sim T \left({g_{0}^2 T^{(4-D)} T^{(D-2)} \over (g_{0} T)^2}
\right)^p (g_0 T)^{(D-1)}
= T^D \ g_{0}^{(D-1)} \, ,
\end{equation}
while the magnetic ($G$ type) contribution is
\begin{equation}
\sim T \left({g_{0}^2 T^{(4-D)} T (g_0 T)^{(D-3)} \over
(g_{0} T)^2} \right)^p
(g_0 T)^{(D-1)} = T^D \ g_{0}^{(D-1)+(D-3)p} \, .
\end{equation}
The electric contribution is  plasmon like for all $p$ and for
$D < 5$.
When $D >3$, the magnetic
contribution is plasmon-like only for $p < (5-D)/(D-3)$.
Also since $p \ge 2$,
this implies $ D < 11/3 $. Thus for $3 < D < 11/3$, only the finite
number of terms, $2 \le p < (5-D)/(D-3)$, give a  plasmon-like
contribution in the magnetic sector. The magnetic contribution
might  also be plasmon-like  for $D \le 3$.
On the other hand, it is easy to see from the equations that
for $D>3$, the magnetic contribution is always subleading to the
electric contribution. Thus the leading plasmon
contribution for $D>3$ is given by eq.(\ref{P31}) of the main
text. The integral may be evaluated by the formulae listed
in Note(1) of this appendix : The second term in (\ref{P31})
drops in dimensional
regularisation while the logarithm  is integrated by considering
first its derivative with respect to $m_{el}^2$. It is amusing
to note
that the peculiar ratio $11/3$ appearing in the above analysis
occurs
in a natural but apparently  unrelated way also in the beta
function. \\

\noindent{4. Beta Function}\\
The beta function is easiest to calculate by using background
field techniques
\cite{Dew}. One first computes (see Abbott \cite{Dew} for
details and
further references)
the wavefunction renormalisation factor $Z_{B}$ of the background
 field
$B_{\mu}$  obtained from its self-energy. In the background
Feynman gauge one can obtain
\begin{equation}
\Pi^{(B)}_{(\mu \nu),(ab)}(K) = -g^2 N_{c} \delta_{ab}
(K^2 g_{\mu \nu} - K_{\mu}
K_{\nu}) \ {(7D-6) \over (2D-2)} \ \mbox{$\displaystyle{\int{d^{D}q
\over  (2\pi)^{D}}}$}  {1 \over Q^2 (Q+K)^2} \,
. \label{SBG}
\end{equation}
Like the usual self-energy (\ref{Se}), this self-energy
(\ref{SBG}) is not
gauge-invariant. Using formulae given in the beginning of this
appendix one may
check that this background field self energy also diverges as
$D \to 2$.
The gauge-invariant information in (\ref{SBG}) comes from the
residue,
$Z_{B}^{(1)}$, of the $\epsilon = (4-D)/2$
pole in $Z_{B}$, when the integrals are computed in dimensional
regularisation. The beta function is given by $\beta(g) =
-{1 \over 2} g^2
{ \partial Z_{B}^{(1)} \over \partial g}$. The $D$ dependent
terms outside
the integrals in (\ref{SBG}) give the famous $11/3$ factor.

\newpage
\baselineskip = 0.5\baselineskip

\end{document}